\documentclass{article}
\usepackage[top=1in,bottom=1in,left=1in,right=1in]{geometry}
\usepackage[utf8]{inputenc}
\usepackage{graphicx}
\usepackage{natbib}
\usepackage{paralist}
\usepackage{amssymb}

\title{Probing the Epoch of Reionization with  a JWST-Wide Field Survey}

\begin{document}

{\raggedright
\Large
Astro2020 Science White Paper \linebreak

{\bf JWST: Probing the Epoch of Reionization with  a Wide Field Time-Domain Survey}\linebreak

\normalsize

\noindent \textbf{Thematic Areas:} 4.Formation  and  Evolution  of  Compact  Objects\\
\textbf{Secondary Thematic Areas:} 7. Cosmology and Fundamental Physics \linebreak

\textbf{Principal Author:}

\date{March 2019}
Name: Lifan Wang
\linebreak						
Institution: Texas A\&M University
\linebreak
Email: lifan@tamu.edu
\linebreak
Phone: 979 318 0388
\linebreak

\noindent
\textbf{Co-authors:} 
\linebreak
J.~Mould (Swinburne), D.~Baade (ESO), E.~Baron (Oklahoma), V.~Bromm (UT, Austin), J.~Cooke (Swinburne), X.~Fan (UA), R.~Foley (UCSC), A.~Fruchter (STScI), A.~Gal-Yam (Weizmann), A.~Heger (Monash), P.~H\"oflich (FSU), D.~A. Howell (UCSB, LCO),  A.~Kashlinsky (NASA GSFC), A.~Koekemoer (STScI), J.~Mather (NASA GSFC), P.~Mazzali (Liverpool), F.~Pacucci (Kapteyn, Yale), F.~Patat (ESO), E.~Pian (INAF OAS), S.~Perlmutter (LBNL), A.~Rest (STScI), D.~Rubin (STScI), D.~Sand (UA), C. Stubbs (Harvard), N.~Suntzeff (TAMU), X.-F.~Wang (Tsinghua), D.~Whalen (Portsmouth), J.~C. Wheeler (UT Austin), and B.~Yue (NAOC)
\bigskip

\noindent

{\bf Abstract}
A stated scientific goal of JWST is to probe the epoch of re-ionization of the Universe at redshifts above 6, to 20 and beyond. At these redshifts, galaxies are just beginning to form and the observable objects are early black holes, supernovae, and cosmic infrared background. The JWST has the necessary sensitivity to observe these targets individually, but a public deep and wide science enabling survey in the wavelength range from 2-5 $\mu$m will be needed to discover these black holes and supernovae, and to cover the area large enough for cosmic infrared background to be reliably studied. This enabling survey will find a large number of other transients and enable supernova cosmology up to z $\sim$ 5, star formation history at high redshift through supernova explosions, faint stellar objects in the Milky Way, and galaxy evolution up to  z approaching  10. The results of this survey will also serve as an invaluable target feeder for the upcoming era of ELT and SKA.
}

\bigskip

\noindent

\section{JWST and the Epoch of Reionization}
Just as a famous early goal of the Hubble Space Telescope (HST) was to measure the Hubble Constant, so a well publicized early goal of the James Webb Space Telescope (JWST) is to find and characterize the first stars in the Universe. Population III (Pop III) does not come with a handbook of how to do this. In a ‘shallow’ survey, JWST takes us into a discovery space of transient phenomena at AB$\sim$27 mag of which we have no knowledge, and create discoveries of early black holes and supernovae (SNe). The world needs to know about transient phenomena in the very early Universe, as transients are prompt tracers of the constituents of the Universe. This knowledge may enable new sciences for the ELT and provide essential data for science goals of WFIRST.

Supernovae (SNe) are a defining characteristic of Pop III. When the first SNe occur, Pop III ends, as they inject first metals into the interstellar medium. Inter alia the breakthrough science with JWST, we can recognize the first stellar explosion as they go SNe. If this can be achieved, a wealth of follow up science opens up for the community, not wholly predictable at the time of launch, but possibly among the significant outcomes of the mission. 

\section{Outline of a JWST Enabling Science Survey}

\subsection{The scientific objectives}
Instead of focusing on programs targeting specific science objectives, one may consider a JWST Enabling Science Survey (JESS) program that can obtain the baseline data required by several major probes of the epoch of re-ionization. 

{\bf Pop III supernovae:} A primary goal of a JESS may be Pop III supernovae. 
High-redshift SNe can directly probe the first generations of stars and their formation
rates because they can be observed at great distances and, to some degree, the mass of
the progenitor can be inferred from the light curve of the explosion \citep{Tanaka:2013,deSouza:2013MNRAS.436.1555D,DeSouza:2014MNRAS.442.1640D,Wang:2017}. At $\sim$  100 M$_\odot$ non-rotating Pop III stars can encounter the pair instability \citep{Heger:2003}.  At 100–140 M$_\odot$ the pair instability causes the ejection of multiple, massive shells instead of the complete destruction of the star \citep{Woosley:2017ApJ...836..244W}. 
Rotation which can cause stars to explode as PI SNe at masses as low as $\sim$ 85 M$_\odot$ \citep{Chatzopoulos:2015ApJ...799...18C}. Above 260 M$_\odot$ stars encounter photo disintegration with 140–260 M$_\odot$ stars exploding as highly energetic thermonuclear PI SNe \citep{Heger:2002}. 
The rate of Pop III SNe is recently estimated by \citep{Moriya:2019} who show that with a 2.1 $\mu$m survey down to 26.5 mag, one may find about 8 Pop III SNe over a field of 1 sq degree in a 5 year survey at $z$ above 5. These first generation SNe may have an extremely extended light curve that lasts for about 10 years. JESS may survey a comparable sky area down to much deeper magnitudes and redder wavelength than LSST or Subaru can, thus allowing the Pop III supernovae, if they exist at all, to be studied with exquisite details. 

{\bf Direct collapse black holes:} The first SMBH seeds formed when the Universe was younger than $\sim$ 500 Myr and played an important role in the growth of early (z $\sim$ 7) SMBHs \citep[e.g.,][]{Pacucci:2015,Natarajan:2017}. 
Such events could occur in atomically cooled halos at z $\sim$ 15–20, heralding the birth of Direct Collapse Black Holes (DCBHs), and thus the first quasars. 
Much progress has been made on DCBHs in recent years in understanding their formation, growth and observational signatures, but many questions remain unanswered, and we are yet to detect these sources. \cite{Natarajan:2017} predicted  the observational properties and JWST detectability of black hole seeds, formed by the direct collapse of high-redshift halos \citep[e.g.,][]{Bromm:2003}, or as remnants of Pop III stars \citep[e.g.,][]{Volonteri:2003}. These black holes are bright enough that JWST can easily observe them up to z beyond 10 \citep{Wang:2017}. Their comparatively lower mass also makes them more prone to short term variability than SMBHs. This makes them a prime target for a wide field transient survey. However, as the compact objects diminish at higher and higher redshifts, only a wide field survey can promise the discovery of the earliest DCBHs, and provide the clear links to many of the mysteries related to reionization and galaxy formation. 

{\bf Superluminous supernovae: } SLSNe \citep[e.g.][]{Gal-Yam:2018arXiv181201428G} can be another critical probe of the deaths of the first stars, because those events are bright (M$\, <\,-$20.5 mag) \citep{DeCia:2018ApJ...860..100D,Lunnan:2018ApJ...852...81L}, and have UV-luminous SEDs in the rest-frame. SLSNe are typically 2–3 magnitudes brighter than SNe Ia, and 4 or more magnitudes brighter than common core collapse (CC) SNe. Observationally, there are two known types of SLSNe. Hydrogen-rich SLSN II are thought to gain their extremely high luminosity from the collision between the expanding SN ejecta and a dense circumstellar shell ejected before the explosion. The mechanism of the explosion itself and their relation to Pop III SNe are less clear. Hydrogen-deficient SLSNe, called SLSN I, do not show signs for such a violent collisional interaction; thus, they are thought to be powered by the spin-down of a magnetar resulting from the core collapse of an extremely massive progenitor star. These stars are discovered at low mass galaxies and are very likely to be associated with a low-metallicity environment. They are bright enough that they can be observed to z beyond 20. A major challenge will be to find them. No ground-based facilties are capable of discovering them at z beyond 6. They are expected to be very rare and only a deep survey down to 27th mag in the IR with JWST, with a field of view of about 1 square degree, can such supernovae be discovered \citep{Wang:2017,Regos:2018arXiv181100891R}. 

{\bf The first thermonuclear Type Ia supernovae:} Whether there are any thermonuclear explosions of white dwarfs at z at 4--6 is still an open question. The presence of Type Ia SNe above z $>$ 2 requires the existence of a prompt channel for their progenitors which has negligible delay time after the formation of the carbon-oxygen (C/O) white dwarf. A JESS survey will discover thermonuclear SNe out to redshifts of 6 if they exist at all. These are the first generation of thermonuclear white dwarf supernovae that enriches the Universe with a significant amount of iron and can be critical in the chemical evolution of the Universe. The thermonuclear SNe are from a population of progenitors with well constrained ages and metallicity and can thus provide critical information on the physics related to the thermonuclear explosion. More details can be found in \cite{Wang:2017}.  

{\bf The Hubble diagram up to z of 5-6:} Based on our best estimate of the rates of SNeIa, the JESS survey can find enough SNe out to redshifts of 6 to build a Hubble diagram out to that redshift. These data are essential in constraining cosmological parameters as well as enabling robust systematic controls related to the evolution of SN properties with redshift. The systematic control of SN properties is of fundamental importance to cosmological studies with WFIRST. 
 Type Ia SNe at such redshifts
will experience time dilation by a factor of 4–5. Once discovered, they are expected to be
visible by JWST for about 6 months, and bright enough for photometric and spectroscopic
followup. Some critical followups can be done in parallel mode with minimal cost in additional observing time. They can be studied by separate programs aiming to directly measure the cosmic deceleration at those redshifts, the ISM towards these SNe, and stellar evolution at very high
redshifts. The JESS 
program will serve as target feeder to programs aiming to measure precision cosmological
parameters using SNeIa.

{\bf Cosmic Infrared Background: } 
\cite{Kashlinsky:2015ApJ...804...99K} present new methodology to use the cosmic infrared background (CIB) fluctuations to probe sources at 10 $<$ z $<$ 30 from a JWST/NIRCam configuration that will isolate known galaxies to 28 AB mag at 0.5-5 $\mu$m. At present significant mutually consistent source-subtracted CIB fluctuations have been identified in the Spitzer and AKARI data at $\sim$2-5 $\mu$m. They evaluate CIB contributions from remaining galaxies and show that the bulk of the high-z sources will be in the confusion noise of the NIRCam beam, requiring CIB studies. The accurate measurement of the angular spectrum of the fluctuations and probing the dependence of its clustering component on the remaining shot noise power would discriminate between the various currently proposed models for their origin and examine the flux distribution of its sources. They show that the contribution to CIB fluctuations from remaining galaxies is large at visible wavelengths for the current instruments precluding detection of the putative Lyman-break of the CIB fluctuations. They demonstrate that with their proposed JWST configuration such measurements will enable probing the Lyman-break. They develop a Lyman-break tomography method to use the NIRCam wavelength coverage to identify or constrain, via the adjacent two-band subtraction, the history of emissions over 10 $<$ z $<$ 30 as the universe comes out of the “Dark Ages.”
Multi-color surveys across an area larger than 1 square degree is needed for CIB measurement. 

{\bf Synergies with LSST and WFIRST: } The Lyman break moves beyond LSST sensitivity window at z $\gtrsim$ 10. This makes LSST useful in eliminating foreground objects from the JESS survey. JESS enables wide field astronomy at z above 10. With its 1 square degree multi-color and slitless grism coverage in the ecliptic poles, JESS is powerful enough to positively affect the selection of survey areas for cosmological studies using  the future WFIRST mission.

\subsection{A strawman JESS design}

The design of such a survey must be a community effort. What are provided here are only preliminary assessments of the survey field, cadence, and feasibility of the investigation.

{\bf Survey field: }
A community JWST survey field was proposed by \cite{Jansen:2018PASP..130l4001J}  which has   a $\sim$ 14 arcmin field located within JWST's northern continuous viewing zone (CVZ). The JESS field needs to be located in the CVZs to allow for comprehensive time-domain coverage. A southern field is important as it enables follow-up observations with the powerful ESO ELT and SKA in the future. The southern CVZ field is perhaps too close to the LMC for CIB studies, but the transient sciences are not significantly affected. 
A southern JESS survey area can be split into two subareas straddling the CVZ.

{\bf Cadence: } Due to the very large range of time dilation, the survey cadence can have a broad range. We envision a two-cadence survey with varying survey field size. The core survey field has a cadence of 3 months, and a survey field of about 0.1-0.2 square degrees. A broader survey field with a cadence of 6 months and a survey area of 0.5 square degrees. 

{\bf Survey size: } The transient survey field can move around with the overlapping area having more frequent time coverage to meet the observational requirement for shorter term transients. This can naturally lead to an increased total survey area, perhaps to as large as 1.63 square degrees from a preliminary design study we have made.

{\bf Survey filters and depth: } The primary transient survey field will reach 27th mag in the filters F150W2 and F322W2 with a S/N of 10 and in filters F200W and F444W with an S/N higher than 4 at any single epoch. More color coverage may be needed for CIB studies.

{\bf Required survey time:} A JESS of a size of $\sim$ 1 sq degree will need about 450 hours per year, with coverage in four survey filters of depths given above. A 1 square degree, 180 day cadence survey will require about 220 hours. A 0.2 square degree, 90 day cadence survey will cost less than 200 hours. The survey design will need to accommodate as many science cases as possible. The numbers provided here serve only as a crude estimate. 

{\bf Transients follow-up observations: } There can be three modes of transient follow-ups for JESS. First, many transients can be studied during the survey with coordinated parallel observations. Second, a small amount of follow-up time may be necessary for the survey program to secure data for high priority targets. The third mode could be independent proposals from the community for follow-ups of transients in different scientific categories. In all these observing programs, careful planning of parallel observations will be critically important. For example, during the NIRCAM imaging survey, NIRISS can be used to acquire slitless spectra for a sub-field within the JESS field, which may provide essential data for a broad range of transient targets. Even deeper NIRCAM images within the JESS field can also be obtained if targeted NIRSPEC follow-ups of very faint objects are planned, this could be particularly useful if high S/N ratio data are needed such as is the case of supernova cosmology. 

\subsection{Community access to raw and processed data: } The data can be made public as soon as specific quality controls are met. Teams working on the data analysis, especially on  transient detection, should release all discoveries to the public as fast as possible. This should enable proposals to be submitted for follow-up observations of transients discovered by JESS.

Deep coordinated surveys from the radio to the X-rays can be arranged for these fields by other observatories, thus defining a truly well studied regions for time-domain astronomy.

In summary, a coordinated JWST Enabling Science Survey that requires a significant amount of investment of observing time can be critical in achieving one of the prime science objectives of JWST: the physics of the formation of first compact structures in the Universe. The science capacity of JESS is too big and the scientific scope is too broad for any single collaborations. NASA/STScI may provide guidance or serve to coordinate such efforts.


\end{document}